\documentclass[letterpaper,12pt]{article}
\usepackage{tabularx} 
\usepackage{amsmath,amssymb}  
\usepackage{graphicx} 
\usepackage[margin=1in,letterpaper]{geometry} 
\usepackage{cite} 
\usepackage[final]{hyperref} 
\hypersetup{
	colorlinks=true,       
	linkcolor=blue,        
	citecolor=blue,        
	filecolor=magenta,     
	urlcolor=blue         
}
\usepackage{blindtext}
\usepackage{graphicx}
\graphicspath{ {./figures/} }
\usepackage{float}
\usepackage{tensor}

\DeclareMathOperator{\argcosh}{argcosh}

\begin{document}

\begin{titlepage}

\begin{center}
\hfill TU-1256\\
\hfill KEK-QUP-2025-0005
\vskip 1.in

\renewcommand{\thefootnote}{\fnsymbol{footnote}}

{\Large \bf
Inflation with Gauss-Bonnet correction:
\\ \vspace{2.5mm}
beyond slow-roll
}

\vskip .5in

{\large
Kamil Mudruňka$^{(a)}$\footnote{mudrunka.kamil.q5@dc.tohoku.ac.jp}
and
Kazunori Nakayama$^{(a,b)}$\footnote{kazunori.nakayama.d3@tohoku.ac.jp}
}

\vskip 0.5in

$^{(a)}${\em 
Department of Physics, Tohoku University, Sendai 980-8578, Japan
}

\vskip 0.2in

$^{(b)}${\em 
International Center for Quantum-field Measurement Systems for Studies of the Universe and Particles (QUP), KEK, 1-1 Oho, Tsukuba, Ibaraki 305-0801, Japan
}

\end{center}
\vskip .5in

\begin{abstract}

If a coupling between the inflaton and the Gauss-Bonnet term is introduced, many models of inflation that were ruled out by the most recent Planck data can be made viable again. The predictions for the scalar spectral index and tensor-to-scalar ratio are typically computed using the slow-roll approximation. In this paper we instead study the full equations of motion and determine the necessary initial conditions for reasonable inflation epoch. We derive the conditions under which the Friedmann equation admits positive solutions for the Hubble parameter. Then we study the possibility of the inflaton becoming trapped in a local potential minimum induced by the Gauss-Bonnet term. Finally we demonstrate the results on monomial potential models with a quadratic and a quartic potential and show that the slow-roll approximation becomes imprecise in the quartic case.

\end{abstract}

\end{titlepage}

\tableofcontents

\renewcommand{\thefootnote}{\arabic{footnote}}
\setcounter{footnote}{0}

\section{Introduction}

Cosmological inflation is nowadays the most widely accepted solution to the problems of old Big-Bang cosmology. However, with many candidate models in existence, the precise mechanism behind it remains subject to discussion. The latest cosmic microwave background (CMB) temperature fluctuation measurements by the Planck satellite~\cite{Planck:2018jri} provided us with valuable data which ruled out many of the known inflation models. These include the simple monomial potential models~\cite{Linde:1983gd}. 

Among other ways~\cite{Destri:2007pv,Nakayama:2013jka,Nakayama:2013txa,Kallosh:2013hoa,Kallosh:2013yoa,Nakayama:2014hga,Galante:2014ifa}, one possible way to make the monomial potential models viable again is to consider higher order curvature corrections to general relativity.
Specifically we are going to focus on the Gauss-Bonnet (GB) term correction which is motivated by low energy effective actions of heterotic string theories~\cite{Antoniadis:1993jc}, whose cosmological effects (also including late time cosmology) had been extensively studied~\cite{Kawai:1998bn,Hwang:1999gf,Nojiri:2005vv,Nojiri:2005jg,Carter:2005fu,Satoh:2007gn,Kawai:1998ab,Satoh:2008ck,Guo:2009uk,Guo:2010jr,Jiang:2013gza,Kawai:2017kqt,Chakraborty:2018scm,Odintsov:2018zhw,Odintsov:2019clh,Pozdeeva:2020apf,Pozdeeva:2021iwc,Kawai:2021bye,Kawai:2021edk,Khan:2022odn,Tsujikawa:2022aar,Odintsov:2023aaw,Odintsov:2023weg,Kawai:2023nqs,Mudrunka:2023wxy,Nojiri:2023mvi,Pinto:2024dnm,Nojiri:2010wj,TerenteDiaz:2023iqk,Millano:2023czt,Millano:2023gkt,Yogesh:2024mpa}. Inclusion of the GB term can not only change the prediction for the CMB observables~\cite{Satoh:2008ck,Guo:2010jr} and make the model compatible with the most recent Planck data~\cite{Jiang:2013gza,Pozdeeva:2020apf,Mudrunka:2023wxy}, but also reveals new interesting phenomena like blue tensor perturbation power spectra and a unique footprint in the gravitational wave (GW) background produced during reheating~\cite{Mudrunka:2023wxy}.

In Ref.~\cite{Mudrunka:2023wxy} we studied the modified prediction of the scalar spectral index and tensor-to-scalar ratio in a simple quadratic inflaton potential with GB correction term. 
This was done using the slow-roll formalism presented in Ref.~\cite{Satoh:2008ck}. In this paper we verify the validity of the slow-roll approximation and explore phenomena related to initial conditions which may limit the validity of the slow-roll formulas for other models. The results will be demonstrated on a model with a quadratic and quartic potential.

The paper is structured as follows. After introduction of the GB model, its action and field equations and the slow-roll approximation used throughout this paper are summarized in Sec.~\ref{sec:GB}. 
In Sec.~\ref{sec:beyond}, we discuss the errors that might be introduced into the calculations of the CMB observables by the slow-roll approximation. In Sec.~\ref{sec:InitCond}, we discuss two kinds of initial condition problems that exist in the GB inflation model. We derive the necessary and sufficient conditions for the Friedmann equation to admit positive solutions for $H$. Then we examine the existence of a critical initial field values, beyond which the model predicts everlasting inflation unless extremely precisely tine tuned initial conditions are considered. We show that for monomial potentials this problem does not affect the model in the parameter range necessary to make it compatible with the Planck data. 
Then, in Sec.~\ref{sec:mono}, we compute the predictions for the CMB observables for quadratic and quartic potentials using both the slow-roll and full equation of motion approach and discuss the precision of the slow-roll approximation in these two cases. 
We also compute the energy spectrum of the GWs produced by decay of the inflaton into gravitons during reheating caused by the GB term for the quadratic potential.

\section{Slow-roll inflation with Gauss-Bonnet correction}
\label{sec:GB}

We consider the GB-coupled inflation model based on the action
\begin{gather*}
   S=\int d^4 x \sqrt{-g}\left[\frac{1}{2}M_{\rm pl}^2R - \frac{1}{2}\partial^\mu \phi \partial_\mu \phi-V\left(\phi\right)-\frac{1}{16}\xi\left(\phi\right)R_{\rm GB}^2\right],
\end{gather*}
where $\phi$ is the inflaton, $M_{\rm pl}$ is the reduced Planck scale,
\begin{gather*}
   R_{\rm GB}^2=\tensor{R}{^\mu^\nu^\alpha^\beta}\tensor{R}{_\mu_\nu_\alpha_\beta}-4\tensor{R}{^\mu^\nu}\tensor{R}{_\mu_\nu}+R^2,
\end{gather*}
is the GB term, $V(\phi)$ is the inflaton potential and $\xi(\phi)$ is a coupling function. Due to the presence of this coupling the GB term is no longer a topological invariant and has a nontrivial contribution to the field equations even in $d=4$ spacetime dimensions. 

Taking the variation with respect to the metric yields the modified Einstein field equations~\cite{Nojiri:2005jg}
\begin{equation}
\label{eqTotalGravFieldEq}
M_{\rm pl}^2\tensor{G}{_\mu_\nu}+\frac{1}{4}R\nabla_\mu \nabla_\nu\xi+\frac{1}{2}\tensor{G}{_\mu_\nu}\nabla^2\xi-\left(\nabla^{\mathstrut}_\rho\nabla^{\mathstrut}_{(\mu}\xi\right)\tensor{R}{_{\nu)}^\rho}+\left(\frac{1}{2}\tensor{g}{_\mu_\nu}\tensor{R}{^\rho^\sigma}-
\frac{1}{2}\tensor{R}{_\mu^\rho_\nu^\sigma}\right)\nabla_\rho\nabla_\sigma\xi=\tensor{T}{_\mu_\nu},
\end{equation}
where $G_{\mu\nu}=R_{\mu\nu}-\frac{1}{2}Rg_{\mu\nu}$ is the Einstein tensor and $T_{\mu\nu}$ is the energy-momentum tensor. The Friedmann equations for a spatially flat Friedmann-Lemaitre-Robertson-Walker universe become 
\begin{equation}
\label{eq:Friedmann1}
    3 M_{\rm pl}^2 H^2-\frac{3}{2}H^3\dot{\xi}=\frac{1}{2}\dot{\phi}^2+V\left(\phi\right),
\end{equation}
\begin{equation}
\label{eq:Friedmann2}
    \ddot{\phi}+3H\dot{\phi}+\frac{\partial V}{\partial\phi}+\frac{3}{2}\frac{\partial\xi}{\partial\phi}H^2\left(\dot{H}+H^2\right)=0,
\end{equation}
where $H=\dot a/a$ is the Hubble parameter with $a(t)$ being the cosmic scale factor.

Let us recall the slow-roll approximation presented in Ref.~\cite{Satoh:2008ck}.\footnote{
    The equivalence between the formalism of Ref.~\cite{Satoh:2008ck} and other literature was explicitly checked in Appendix of Ref.~\cite{Mudrunka:2023wxy}.
}
We assume that the contributions of $\dot{H}$ is negligible as the expansion is almost de Sitter and that \eqref{eq:Friedmann1} is heavily dominated by $V$ even in the presence of the GB term. Mathematically we describe these requirements as 
\begin{equation}
\label{eqSlowRollCondition}
\left|\frac{\ddot{\phi}}{H\dot{\phi}}\right|\ll 1,\quad\frac{\dot{\phi}^2}{M_{\rm pl}^2 H^2}\ll 1,\quad \left|\frac{H\dot{\xi}}{M_{\rm pl}^2}\right|\ll 1,\quad \left|\frac{\dot{H}}{H^2}\right|\ll 1.
\end{equation}
The slow-roll Friedmann equations become
\begin{equation}
\label{eqSlowRollFriedmann}
3 M_{\rm pl}^2H^2=V,\quad \dot{\phi}=-\frac{1}{3H}\frac{\partial V}{\partial \phi}-\frac{1}{2}H^3\frac{\partial \xi}{\partial \phi}.
\end{equation}
The field equation can be conveniently recast in terms of e-folds as
\begin{equation}
    \frac{d\phi}{dN}=-\frac{1}{V}\frac{\partial V}{\partial\phi}-\frac{1}{6}V\frac{\partial\xi}{\partial\phi}.
\end{equation}
The consistency of this approximation requires that the 5 parameters
\begin{align}
\epsilon=\frac{M_{\rm pl}^2}{2 V^2}\left(\frac{\partial V}{\partial\phi}\right)^2,
\quad
\eta=\frac{M_{\rm pl}^2}{V}\frac{\partial^2 V}{\partial\phi^2},
\quad
\alpha=\frac{1}{4 M_{\rm pl}^2}\frac{\partial V}{\partial\phi}\frac{\partial \xi}{\partial\phi},
\quad
\beta=\frac{V}{6 M_{\rm pl}^2}\frac{\partial^2 \xi}{\partial\phi^2},
\quad
\gamma=\frac{V^2}{18 M_{\rm pl}^6}\left(\frac{\partial \xi}{\partial\phi}\right)^2.
\label{epsilon}
\end{align}
are very small in magnitude during the inflation epoch.

In this paper we will consider a dilaton-like coupling function 
\begin{equation}
    \xi(\phi)=8\xi_0 e^{-\lambda\phi},~~~~~~\lambda,\xi_0>0
\end{equation}
which may be theoretically motivated by by corrections to gravity from string theories~\cite{Antoniadis:1993jc}. 
We also assume that the inflaton is rolling down the potential starting from $\phi > 0$ toward smaller $\phi$.
Importantly we have 
\begin{equation}
    \label{eq:negativeXiDerivative}
    \frac{\partial\xi}{\partial\phi}<0,
\end{equation}
which means that the GB term will slow down the inflaton field as it rolls down the potential. In order to obtain a reasonably strong correction for a potential of the form $V(\phi)=g\phi^n$ we need $\xi_0\approx\frac{1}{g}$. Lastly we note that for the case \eqref{eq:negativeXiDerivative} we get
\begin{equation}
    \frac{H\dot{\xi}}{\rm M_{pl}^2}\approx-\frac{4}{3}\alpha-\gamma>0,~~~\alpha<0,
\end{equation}
therefore $\gamma<-4/3\alpha$. When discussing the validity of the slow roll approximation it is only necessary to ensure that $\alpha$ and $\beta$ are small.

The general metric perturbation can be decomposed into a scalar, vector and tensor sectors. The scalar sector can be parametrized as
\begin{equation}
ds^2=a\left(\tau\right)^2\left[-\left(1+2A\right)d\tau^2+2\partial_i B dx^i d\tau + \left(\delta_{ij}+2\psi\delta_{ij}+2\partial_i\partial_j E\right)dx^i dx^j\right].
\end{equation}
The linearized version of \eqref{eqTotalGravFieldEq} yields one master equation for the canonical variable $\Psi_k\equiv M_{\rm pl}A_\psi \psi_k$ in the momentum space:
\begin{equation}
\label{eqScalarPertPsiCanonical}
\Psi_k^{\prime\prime}+\left(C^2_\psi k^2-\frac{A_\psi^{\prime\prime}}{A_\psi}\right)\Psi_k=0.
\end{equation}
where the prime denotes derivative with respect to $\tau$ and
\begin{align}
&C_\psi^2=\frac{2}{A^2_\psi}\left[\left(\frac{a^2 X^2}{\mathcal{H}Y}\right)^{\prime}-a^2\left(1-\frac{\ddot{\xi}}{2}\right)\right], \label{Cpsi}\\
&A^2_\psi=6a^2 X\left[1-\left(1-\frac{\rho^2}{6}-\sigma\right)\frac{X}{Y^2}\right], \label{Apsi} \\
&\sigma\equiv\frac{H\dot\xi}{M_{\rm pl}^2},\quad \rho^2\equiv\frac{\dot{\phi}^2}{M_{\rm pl}^2H^2}\mathrm{,}\quad X\equiv1-\frac{\sigma}{2}\mathrm{,}\quad Y\equiv1-\frac{3}{4}\sigma.
\end{align}
This equation can be analytically solved under the approximation that the derivatives of $A_\psi$ and $C_\psi$ are negligible. The general solution in the momentum space is 
\begin{align}
\label{eqHankel1}
	\widetilde\Psi_k (\tau) = e^{\frac{i(2\nu_\psi+1)\pi}{4}} \frac{1}{\sqrt{2C_\psi k}} \sqrt{\frac{-\pi C_\psi k\tau}{2}} H_{\nu_\psi}^{(1)}(-C_\psi k\tau),
\end{align}
where the effective potential $\nu_\psi$ is given by
\begin{align}
	\nu_\psi^2 = \frac{1}{4} + \frac{\tau^2 A_\psi''}{A_\psi} \simeq \frac{9}{4} + 9\epsilon -3\eta- \alpha -3\beta,
\end{align}
After a first order expansion in the five slow-roll parameters this gives the spectral index $n_s$ to be
\begin{equation}
	n_s-1=3-2\nu_\psi= -6\epsilon+2\eta+\frac{2}{3}\alpha+2\beta.  \label{ns}
\end{equation}
The tensor sector of the linearized perturbations can be written as
\begin{equation}
\label{eqTensorPertMetric}
ds^2=a\left(\tau\right)^2\left[-d\tau^2+\left(\delta_{ij}+h_{ij} \right)dx^i dx^j\right],
\end{equation}
which yields the master equation for the two canonical GW polarizations  $h^c_{k,\lambda}\equiv M_{\rm pl}A_T h_{k,\lambda}/2$ in the momentum space
\begin{gather}
{ h^{c\,\prime\prime}_{k,\lambda}}+\left(C^2_T k^2-\frac{A_T^{\prime\prime}}{A_T}\right)h^c_{k,\lambda}=0,
\end{gather}
where
\begin{equation}
   A_T^2\equiv a^2\left(1-\frac{\sigma}{2}\right),~~~C_T^2\equiv \frac{a^2}{A_T^2}\left(1+\frac{\sigma}{2}-\frac{\xi^{\prime\prime}}{2a^2 M_{\rm pl}^2}\right).
\end{equation}
For the case of $\partial\xi/\partial\phi<0$, the speed of sound of the tensor modes becomes larger than 1. This issue is often discussed in the light of the GW170817 event, which constrains $C_T^2=1$ in the late-time Universe \cite{Gangopadhyay:2022vgh,Nojiri:2023mbo,Odintsov:2020sqy,Odintsov:2020zkl}. However, we should note that this observational restriction only applies to the late-time evolution, not to the inflation epoch. Therefore, we do not consider our model to contradict this.

This equation can again be analytically solved under the approximation that the derivatives of $A_T$ and $C_T$ are negligible. The general solution in the momentum space is 
\begin{align}
\label{eqHankel2}
	\widetilde h_{k,\lambda} (\tau) = e^{\frac{i(2\nu_T+1)\pi}{4}} \frac{1}{\sqrt{2C_T k}} \sqrt{\frac{-\pi C_T k\tau}{2}} H_{\nu_T}^{(1)}(-C_T k\tau),
\end{align}
where $\nu_T$ is given by
\begin{align}
	\nu_T^2 = \frac{1}{4} + \frac{\tau^2 A_T''}{A_T} \simeq \frac{9}{4} + 3\epsilon + \alpha.
\end{align}
After a first order expansion in the five slow-roll parameters this gives the tensor spectral index $n_T$ and the tensor-to-scalar ratio $r$ to be
\begin{align}
&n_T=3-2\nu_T=-2\epsilon-\frac{2}{3}\alpha,~~~r\simeq 16\epsilon+\frac{32}{3}\alpha+4\gamma.   \label{nT}
\end{align}

We would also like to draw attention to a new slow roll approximation proposed in \cite{Pozdeeva:2024ihc}, which also incorporates the contribution of the GB term to the Friedmann equation and significantly improves the accuracy of the results. The findings of \cite{Pozdeeva:2024ihc} agree with those of this paper. Although both papers study the monomial potential models, our choice of the GB term coupling is different. 

\section{Beyond slow-roll}
\label{sec:beyond}

There are two ways in which the slow-roll approximation introduces errors into our calculations. Firstly, it affects the time evolution of the homogeneous background value of $\phi$. Secondly, the solutions of the Mukhanov-Sasaki equation given in terms of Hankel functions are only approximate. We will not deal with the second issue in this paper because the Hankel functions formulas only have to be evaluated at the horizon exit, when the slow-roll scenario turned out to be valid for the models discussed here. We will omit any factors of $M_{\rm pl}$ in the equations from now on for better readability.


The evolution of $\phi$ at later times can experience a significant correction from the GB term such that the slow-roll approximation breaks down. The two standard slow-roll parameters $\epsilon$ and $\eta$ are monotonous functions of $\phi$ and decrease as one moves up the inflation potential. For the three parameters related to the GB coupling function, $\alpha$, $\beta$ and $\gamma$, we have 
\begin{equation}
    \lim_{\phi\rightarrow0}\alpha=\lim_{\phi\rightarrow0}\beta=\lim_{\phi\rightarrow0}\gamma=0,
\end{equation}
assuming that 
\begin{equation}
    \lim_{\phi\rightarrow0}V=\lim_{\phi\rightarrow0}\frac{\partial V}{\partial\phi}=0.
\end{equation}
At the same time they vanish for large $\phi$ and therefore their effect gradually grows during the inflation until they peak and then diminish again towards the very end. In order to estimate the error introduced by the slow-roll approximation, we can numerically solve the full equations \eqref{eq:Friedmann1} and \eqref{eq:Friedmann2}.

To achieve this we differentiate \eqref{eq:Friedmann1} and eliminate $\dot{H}$ from \eqref{eq:Friedmann2} using the result. The resulting single equation is
\begin{equation}
\label{eq:FullSingleFriedmann}
    \ddot{\phi}=\frac{9 H^{5} \dot{\phi} \left(\xi^\prime\right)^{2} - 3 H^{4} \dot{\phi}^{2} \xi^{\prime} \xi^{\prime\prime} - 12 H^{4} \xi^{\prime} + 18 H^{2} \dot{\phi}^{2} \xi^{\prime}+ 4 H V^{\prime} \dot{\phi} \xi^{\prime} - 24 H\dot{\phi} - 8 V^{\prime}}{3 H^{4} \left(\xi^{\prime}\right)^{2} - 4 H \dot{\phi} \xi^{\prime} + 8}.
\end{equation}
$H$ is given by \eqref{eq:Friedmann1} where the root is searched for near the positive solution of the case $\xi^{\prime}=0$. The formulas used for this together with a detailed analysis are presented in Sec.~\ref{sec:InitCond}. While the slow-roll approximation allows only for a single solution (assuming that we want an expanding universe), \eqref{eq:Friedmann1} might have an additional positive root based on the sign of the extra term. However, unless a very specific choice of the coupling function $\xi$ is made, as the inflation ends and $\phi$ enters the oscillatory phase and reaches the amplitude of the oscillation where $\dot{\phi}=0$, this root gets shifted to infinity, making it rather nonphysical.

Unlike the slow-roll equation, where only the product $\xi V$ appears, in the case of \eqref{eq:FullSingleFriedmann} we need 3 parameters to describe the model: $\xi_0$ and $\lambda$, which determine the strength and peak effect location of the GB term coupling, and $g$ which determines the (effective) mass of the inflaton field and/or its self-interaction. However, it turns out that, unless the GB term coupling is very strong, in reality once again only the product $g\xi_0$ affects the evolution. This is better seen when the Friedmann equation is parametrized in terms of $N=\int H dt$ rather than $t$. Eq.~\eqref{eq:Friedmann2} becomes
\begin{equation}
    \label{eq:Friedmann2Efolds}
    \ddot{\phi}=\frac{9 H^{6} \frac{d\phi}{dN} \left(\xi^\prime\right)^{2} - 3 H^{6} \left(\frac{d\phi}{dN}\right)^{2} \xi^{\prime} \xi^{\prime\prime} - 12 H^{4} \xi^{\prime} + 18 H^{4} \left(\frac{d\phi}{dN}\right)^{2} \xi^{\prime}+ 4 H^2 V^{\prime} \frac{d\phi}{dN} \xi^{\prime} - 24 H^2\frac{d\phi}{dN} - 8 V^{\prime}}{3 H^{4} \left(\xi^{\prime}\right)^{2} - 4 H^2 \frac{d\phi}{dN} \xi^{\prime} + 8},
\end{equation}
where $\ddot{\phi}$ can be expanded as
\begin{equation}
    \ddot{\phi}=H\frac{d}{dN}\left(H\frac{d\phi}{dN}\right)=H^2\frac{d^2\phi}{dN^2}+\frac{1}{2}\frac{d (H^2)}{dN}\frac{d\phi}{dN},
\end{equation}
and $\eqref{eq:Friedmann1}$ becomes
\begin{equation}
\label{eq:Friedmann1Efolds}
    H^2\left(3-\frac{1}{2}\left(\frac{d\phi}{dN}\right)^2\right)=V+\frac{3}{2}H^4\frac{\partial \xi}{\partial \phi}\frac{d\phi}{dN}.
\end{equation}
Denoting 
\begin{equation}
    P=3-\frac{1}{2}\left(\frac{d\phi}{dN}\right)^2,~~~Q=\frac{3}{2}\frac{\partial\xi}{\partial\phi}\frac{d\phi}{dN},
\end{equation}
we have $P H^2=V+Q H^4$. The solutions are 
\begin{equation}
\label{eq:EnergyFriedmannPertCorrection}
    H^2=\frac{P\pm\sqrt{P^2-4QV}}{2Q}=\frac{PV\left(1\pm\sqrt{1-\frac{4QV}{P^2}}\right)}{2QV}.
\end{equation}
This expression only contains products $V\xi^{\prime}$ and is multiplied by an extra factor of $V$. Looking back at \eqref{eq:Friedmann2Efolds} we see that every time $\xi^{\prime}$ or $\xi^{\prime\prime}$ appears, it has precisely one factor of $H^2$ in front of it. This together with the form of \eqref{eq:EnergyFriedmannPertCorrection} explains the dependency on $\zeta=g\xi_0$ only. This phenomenon can be exploited to speed up the numerical calculations by choosing a large reference value of $g$, which makes the inflation reach the desired amount of e-folds in short cosmological time.

\begin{figure}[H]
\centering
\includegraphics[width=.48\textwidth]{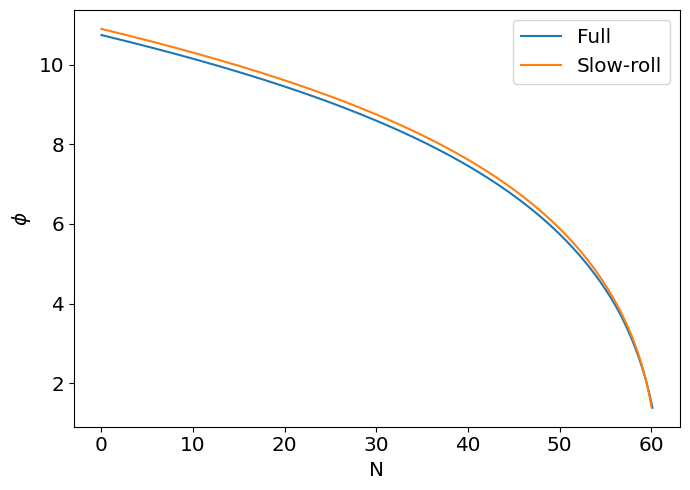}
\includegraphics[width=.48\textwidth]{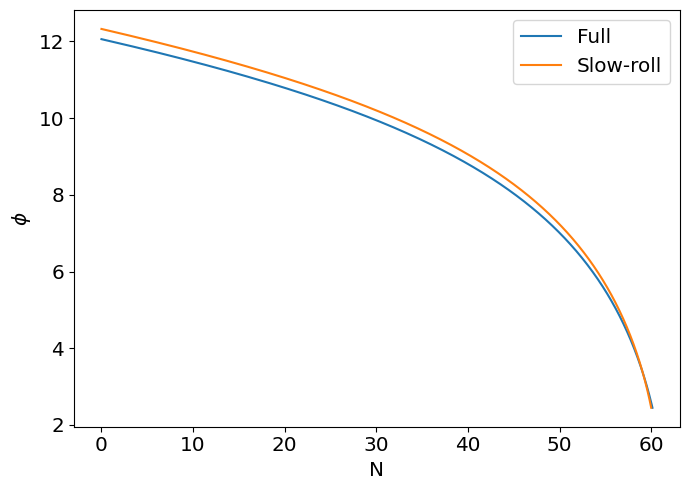}
\includegraphics[width=.48\textwidth]{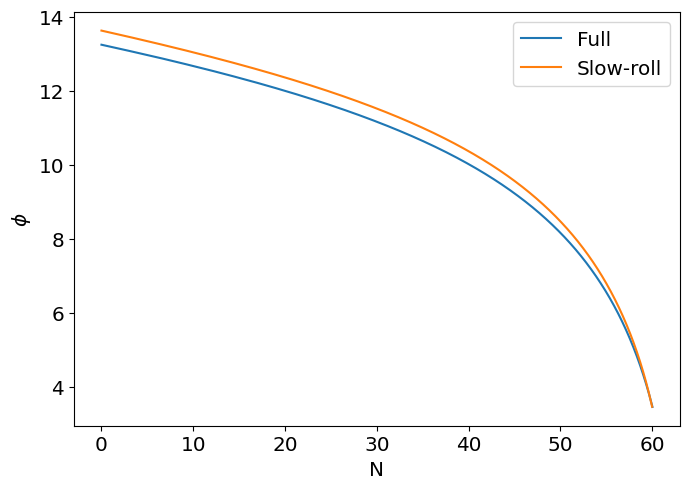}
\caption{Time evolution of $\phi$ for $V=g\phi^2$, $\lambda=0.218$, $\zeta=0.0533$ (top left),
$V=g\phi^3$, $\lambda=0.2913$, $\zeta=0.0124$ (top right),
$V=g\phi^4$, $\lambda=0.345$, $\zeta=0.0022$ (bottom).
The orange plot comes from the slow-roll approximation. The blue plot comes from solving the full equations. In this case, a reference value of $g=2\cdot10^{-12}$  was used.}
\label{figSlowRollVsFull}
\end{figure}

To demonstrate the effect the slow-roll approximation has on the solution, we present graphs of $\phi$ vs $N$ for a model with monomial potential $V=g\phi^n$ for $n=2,3,4$. 
The top left panel of Fig.~\ref{figSlowRollVsFull} shows the case of $n=2$, where we have chosen a reference value of $\lambda=0.218$, $\zeta=0.0533$. The slow-roll approximation predicts a value of $\phi=10.896 M_{\rm pl}$ at the horizon exit for inflation lasting $60$ e-folds. The full equations predicts a value of $\phi=10.743 M_{\rm pl}$. The relative difference between the values is $1.41\%$.
The top right Fig.~\ref{figSlowRollVsFull} shows the case of $n=3$, where we have chosen a reference value of $\lambda=0.2913$, $\zeta=0.0124$. The slow-roll approximation predicts a value of $\phi=12.326 M_{\rm pl}$ at the horizon exit. The full equations predicts a value of $\phi=12.059 M_{\rm pl}$. The relative difference between the values is $2.16\%$.
The bottom panel of Fig.~\ref{figSlowRollVsFull} shows the case of $n=4$, where we have chosen a reference value of $\lambda=0.345$, $\zeta=0.0022$. The slow-roll approximation predicts a value of $\phi= 13.619 M_{\rm pl}$ at the horizon exit. The full equations predicts a value of $\phi=13.238 M_{\rm pl}$. The relative difference between the values is $2.8\%$.

In general the slow-roll approximation overestimates the value of $\phi$ at horizon exit. This is mainly due to the neglected $\frac{3}{2}H^3\dot{\xi}$ term in \eqref{eq:Friedmann1}. For the dilaton coupling this term is positive and increases the overall value of $H$, making the effect of the GB term in \eqref{eq:Friedmann2} stronger while also weakening the effect of $\frac{\partial V}{\partial \phi}$.

\section{Initial conditions for inflation}
\label{sec:InitCond}

\subsection{Conditions for expanding universe}

In the case of zero initial velocity, Eq.~\eqref{eq:Friedmann1} is quadratic in $H$ and always has a positive solution, which allows for inflation. However, given a nonzero initial velocity the equation turns into a cubic one and in general there is no longer anything that would guarantee the existence of a positive solution. 

First let us analyze the situation when the cubic has only one real root. This happens when the discriminant $\Delta$ is negative. Plugging in \eqref{eq:Friedmann1} we get
\begin{equation}
\label{eq:CubicDiscriminant}
    \Delta=-\frac{243}{16}\left(2V+\dot{\phi}^2\right)^2\dot{\phi}^2\left(\xi^{\prime}\right)^2+108V+54\dot{\phi}^2.
\end{equation}
This is a sextic polynomial in $\dot{\phi}$ with a negative coefficient in front of $\left(\dot{\phi}\right)^6$. This implies that for large enough $ \vert\dot{\phi}\vert $ \eqref{eq:Friedmann1} will always have only one real root. To analyze the sign of the solution we transform \eqref{eq:Friedmann1} into a depressed form
\begin{equation}
    t^3+pt+q=0,~~~p=-\frac{4}{3\left(\dot{\phi}\xi^{\prime}\right)^2},~~~q=\frac{9\left(2V+\dot{\phi}^2\right)\left(\dot{\phi}\xi^{\prime}\right)^2-16}{27\left(\dot{\phi}\xi^{\prime}\right)^3},
\end{equation}
where $t=H+\frac{2}{3\dot{\phi}\xi^{\prime}}$. The root of the depressed cubic is given by \cite{Holmes2002TheUO}
\begin{equation}
    t_0=-2\frac{\vert q\vert}{q}\sqrt{-\frac{p}{3}}\cosh\left[\frac{1}{3}\argcosh\left(-\frac{3\vert q \vert}{2p}\sqrt{-\frac{3}{p}}\right)\right].
\end{equation}
This results in an inequality 
\begin{equation}
    \vert t_0 \vert \geq \frac{4}{3\vert \dot{\phi}\xi^{\prime} \vert},
\end{equation}
which implies that the sign of $H$ and $t_0$ is the same. Therefore, positive solution for $H$ exists if and only if $q<0$. For the dilaton-like coupling we have $\xi^\prime<0$. Therefore in the case of a negative $\dot{\phi}<0$ we need
\begin{equation}
    \label{eq:NeqQCOndition}
    9\left(2V+\dot{\phi}^2\right)\left(\dot{\phi}\xi^{\prime}\right)^2-16<0,
\end{equation}
which is satisfied by
\begin{equation}
    \dot{\phi}^2< -V+\frac{\sqrt{9V^2+16\left(\xi^{\prime}\right)^{-2}}}{3}.
\end{equation}
However, the condition that the discriminant \eqref{eq:CubicDiscriminant} is negative demands that
\begin{equation}
\label{eq:discNegative}
    9\left(2V+\dot{\phi}^2\right)\left(\dot{\phi}\xi^{\prime}\right)^2>32,
\end{equation}
which cannot be satisfied simultaneously\footnote{Alternatively this could be satisfied if $V<0$.}. We conclude that in the case of only one real root and negative $\dot{\phi}$, there are no positive solutions for $H$. For positive $\dot{\phi}$, on the other hand, the inequality \eqref{eq:NeqQCOndition} reverses and, combined with the requirement \eqref{eq:discNegative}, it is always automatically satisfied.

In the case of three real roots we can use the formula \cite{Nickalls1993ANA}
\begin{equation}
    t_n=2\sqrt{-\frac{p}{3}}\cos\left[\frac{1}{3}\arccos\left(\frac{3q}{2p}\sqrt{-\frac{3}{p}}\right)+\frac{2\pi n}{3}\right],~~~n=0,1,2.
\end{equation}
This leads to the opposite inequality
\begin{equation}
    \vert t_0 \vert \leq \frac{4}{3\vert \dot{\phi}\xi^{\prime} \vert},
\end{equation}
meaning that it is possible to have both positive and negative solutions for $H$. For $\dot{\phi} > 0$ all the solutions will be negative if and only if all $t_n<\frac{1}{2\vert\dot{\phi}\xi^{\prime}\vert}$, in other words
\begin{equation}
    \cos\left[\frac{1}{3}\arccos\left(\frac{3q}{2p}\sqrt{-\frac{3}{p}}\right)+\frac{2\pi n}{3}\right]<\frac{1}{2}.
\end{equation}
This is, however, impossible to satisfy for all $n=0,1,2$ at the same time. Similarly in case of $\dot{\phi}<0$ there will always be at least one positive root. The outcomes of the various scenarios are summarized in Table~\ref{tablePosNegH}.

\begin{table}[t]
    \centering
\begin{tabular}{ |p{2cm}|p{5cm}|p{5cm}|  }
\hline
 & One real root & Three real roots \\
\hline
$\dot{\phi}>0$ & Positive solution exists &  Positive solution exists \\
\hline
$\dot{\phi}<0$ & No positive solutions &  Positive solution exists \\
\hline
Condition & $9\left(2V+\dot{\phi}^2\right)\left(\dot{\phi}\xi^{\prime}\right)^2>32$ &  $9\left(2V+\dot{\phi}^2\right)\left(\dot{\phi}\xi^{\prime}\right)^2<32$ \\
\hline
\end{tabular}
    \caption{Summary of the existence of positive solutions for $H$ for the cases of one real root and three real roots and both signs of $\dot{\phi}$.}
    \label{tablePosNegH}
\end{table}

We can conclude that classically inflation can happen only if $\dot{\phi}>0$ or 
\begin{equation}
    \label{eq:MaxPhiDot}
    \dot{\phi}^2< -V+\frac{\sqrt{9V^2+32\left(\xi^{\prime}\right)^{-2}}}{3}.
\end{equation}
To check the consistency with the slow-roll approximation we can write
\begin{equation}
    \dot{\phi}^2=\frac{1}{3V}\left(\frac{\partial V}{\partial \phi}\right)^2+\frac{1}{9}V\frac{\partial V}{\partial \phi}\frac{\partial\xi}{\partial\phi}+\frac{1}{4}V^3\frac{\partial\xi}{\partial\phi}=\left(\frac{2}{3}\epsilon+\frac{4}{9}\alpha+\frac{9}{2}\gamma\right)V.
\end{equation}
As long as $\epsilon, \alpha,\gamma \ll 1$ the slow-roll predictions are consistent with the initial condition constraint obtained in this section. 

If $H$ evolves continuously it must pass through $H=0$, which implies $V=\dot{\phi}=0$. Therefore in the classical regime inflation can never happen in such a scenario. Initial conditions heavily dominated by the kinetic term with a positive inflaton field velocity are ruled out by this. 


Eq.~\eqref{eq:MaxPhiDot} also limits the maximum velocity $\phi$ can ever reach during inflation. As $\xi^{\prime}\rightarrow0$ this velocity limit goes to infinity. On the other end we have $\dot{\phi}^2_{\rm max}\rightarrow0$ as $\xi^{\prime}\rightarrow\pm\infty$. This directly limits the total energy available for decay into Standard Model particles and gravitons during reheating. Assuming $V=0$, $\phi=0$ to be the minimum of the inflaton potential, the energy density at the start of the first oscillation is limited to be 
\begin{equation}
    \rho_{\phi}^{\rm max}=\frac{1}{2}\dot{\phi}^2=\frac{2\sqrt{2}}{3\vert\xi^{\prime}\vert}=\frac{2\sqrt{2}}{24\xi_0\lambda},
\end{equation}
where the last equality holds for $\xi=8\xi_0e^{-\lambda\phi}$. 
This translates to a possible maximum temperature of the radiation dominated universe,
\begin{equation}
    T_{\rm max}=\left(\frac{30}{\pi^2 g_*}\frac{2\sqrt{2}}{24\xi_0\lambda\sigma}\right)^{1/4},
\end{equation}
where $g_*$ is the relativistic degrees of freedom.
This result is independent of the inflaton potential\footnote{In reality the dependence on $V$ exists indirectly through the fact that in order to make the effect of the GB term reasonably strong, we need $\xi\approx V^{-1}$.}.

\subsection{Conditions to avoid trapping at local minimum}
\label{sec:Noneternal}

With the presence of the GB term a new phenomenon arises. Looking at \eqref{eq:Friedmann2}, one may define the effective potential as
\begin{equation}
        \label{eq:FullEffPotential}
        \frac{\partial V_{\rm eff}}{\partial\phi}=\frac{\partial V}{\partial\phi}+\frac{3}{2}\frac{\partial\xi}{\partial\phi}H^2\left(\dot{H}+H^2\right).
\end{equation}
We see that for the dilaton-like coupling, where $\partial\xi/\partial\phi<0$, the derivative may become negative at certain values of $\phi$ and $\dot{\phi}$. This would drive the inflaton field to climb up the potential instead of rolling down and it may be trapped at certain minimum of the effective potential.\footnote{
    Even in that case it is possible that the inflaton escapes from the local minimum by quantum tunneling. We do not consider such a case in this paper.
} 

First let us analyze the situation in the slow-roll approximation. For simplicity we will consider a monomial potential model of the form $V=g\phi^n$. The slow-roll effective potential becomes
\begin{equation}
\label{VeffDiffPoly}
    \frac{\partial V_{\rm eff}}{\partial\phi}=gn\phi^{n-1}-\frac{4\lambda\xi_0g^2}{3}\phi^{2n}e^{-\lambda\phi}.
\end{equation}
The zeros of \eqref{VeffDiffPoly} are given in terms of the two real branches of the Lambert W function $W_0$ and $W_1$ as
\begin{equation}
    \label{PhiCritPoly}
    \phi_{\rm crit}=-\frac{n+1}{\lambda}W_{0,1}\left[-\frac{\lambda}{n+1}\left(\frac{3n}{4g\lambda\xi_0}\right)^{1/(n+1)}\right].
\end{equation}
Right-hand side of \eqref{PhiCritPoly} only makes sense when the argument of the Lambert W function is greater than $-e^{-1}$, which serves as a necessary condition for the existence of an ultra-slow-roll phase. 
The effective potential satisfies $\partial V_{\rm eff}/\partial\phi>0$ everywhere if and only if
\begin{equation}
\label{PhiCritPolyCondition}
    \frac{3n}{4g\xi_0}>\frac{1}{\lambda^n}\left(\frac{n+1}{e}\right)^{n+1},
\end{equation}
and if this condition is satisfied then the location of the two zeros $\phi_{\rm crit0}$ and $\phi_{\rm crit1}$ always satisfies
\begin{equation}
    \label{eqUSRCritPoly}
    \phi_{\rm crit0} \leq \frac{n+1}{\lambda}\leq\phi_{\rm crit1}.
\end{equation}

In the case both sides of \eqref{PhiCritPolyCondition} are equal there is an inflection point in $V_{\rm eff}$ and we could expect a period of ultra-slow-roll inflation.\footnote{
    Ultra-slow-roll regime in the GB model with some function $V$ and $\xi$ and associated primordial black hole formation has been discussed in Ref.~\cite{Kawai:2021edk}.
}
In case the inequality is reversed and $\phi$ starts above the smaller of the zeros it will get trapped in the local potential well and the model will lead to an everlasting inflation stabilizing at constant expansion rate. To see that this is indeed the case we can compute $\frac{\partial^2 V_{\rm eff}}{\partial\phi^2}$. Plugging in the fixed points, we have

\begin{equation}
    \frac{\partial^2 V_{\rm eff}}{\partial\phi^2}=\frac{\lambda^2n}{n+1}\left(\frac{1}{W(\theta)}+\frac{1}{W(\theta)^2}\right),~~~\theta=-\frac{\lambda}{n+1}\left(\frac{3n}{4g\lambda\xi_0}\right)^{1/(n+1)}.
\end{equation}
By comparing with the values of the two branches of the $W$ function we see that indeed the fixed point corresponding to the smaller of the two $\phi_{\rm crit}$ is unstable while the other one is stable.

\begin{figure}[h]
\centering
\includegraphics[width=.6\textwidth]{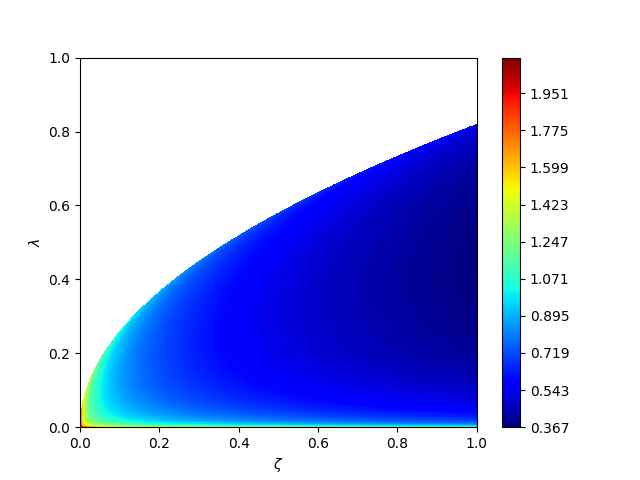}
\includegraphics[width=.6\textwidth]{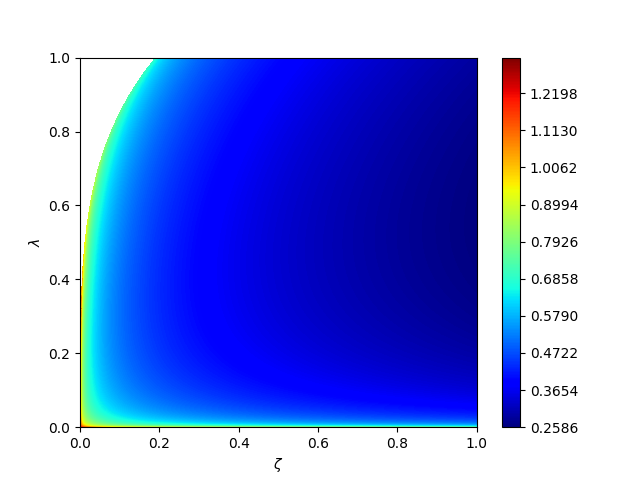}
\caption{Contour plot of $\log_{10}\phi_{\rm crit 0}$ for $n=2$ (top) and $n=4$ (bottom). The white color marks the area where $\frac{\partial V_{\rm eff}}{\partial\phi}>0$ everywhere.}
\label{figCritLow2}
\end{figure}

As a result, models with high values of $\lambda$ are more prone to getting $\phi$ trapped inside a local potential well. This raises a possible problem with the slow-roll calculations. If we pick a value of $\phi$ marking the end of inflation (e.g. $\epsilon=1$) and integrate the first order slow-roll field equation from there, we will never run into the region where the field climbs up. In the cases where $\phi_{\rm crit}$ is small, the effective potential would be extremely flat and $\phi$ would remain almost constant during the first few e-folds after the pivot scale perturbations exit the horizon. The values of $\phi_{\rm crit 0}$ for $n=2$ and $n=4$ models are shown in Fig.~\ref{figCritLow2}.

This leads to a fine tuning problem for models with high $\lambda$ and high $\xi_0$. If the initial field value was just slightly higher, the field would climb up and the inflation would never stop. If it was slightly lower, the inflation would end too soon to reach a total of 60 e-folds. The range of possible initial values of $\phi_{\rm init}$ would be extremely narrow. 

\begin{figure}[H]
\centering
\includegraphics[width=.6\textwidth]{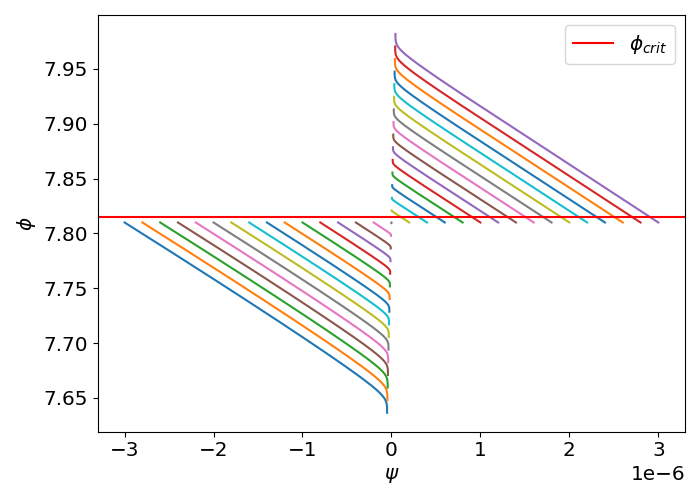}
\includegraphics[width=.6\textwidth]{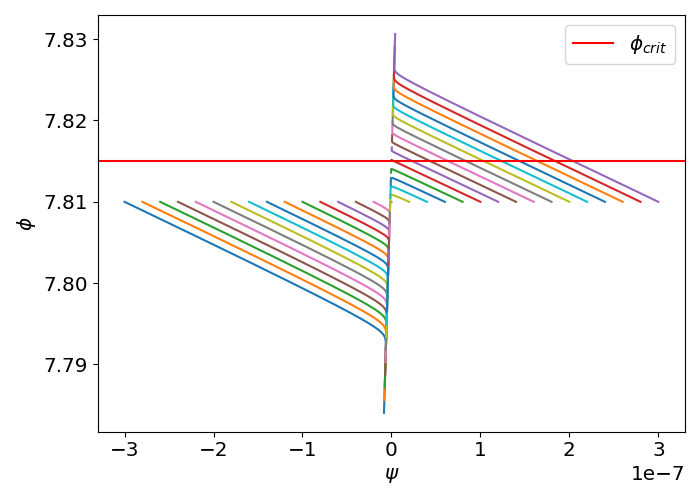}
\caption{Evolution of $\phi$ around $\phi_{\rm crit}$ for 3 e-folds (top) and 10 e-folds (bottom) in the $\psi$,$\phi$ plane. The initial value of $\phi$ is 7.81. We have taken $\lambda=0.2,~~~g=2\cdot10^{-12},~~~\xi_0=0.1\cdot\frac{3}{10}\cdot10^{12}$. The red line marks $\phi_{\rm crit}$.}
\label{figCritlineLowLarge}
\end{figure}

To illustrate what happens near $\phi_{\rm crit}$, we will take a model with $V=g\phi^2$ and choose the parameters to be $\lambda=0.2,~~~g=2\cdot10^{-12},~~~\xi_0=0.1\cdot\frac{3}{10}\cdot10^{12}$.
Evolution of $\phi$ around $\phi_{\rm crit}$ in the plane of $(\phi, \psi)$, where $\psi=\dot{\phi}$, are shown in Fig.~\ref{figCritlineLowLarge}.
In this case the two critical points are $\phi_{\rm crit0}=7.815$ and $\phi_{\rm crit1}=25.631$. 

\begin{figure}[H]
\centering
\includegraphics[width=.7\textwidth]{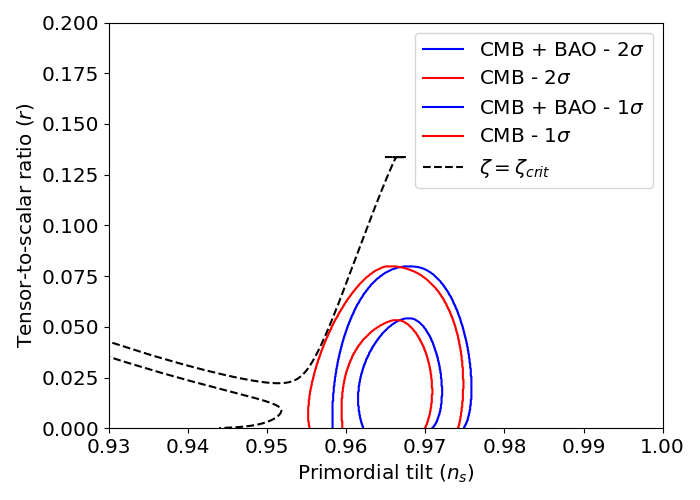}
\includegraphics[width=.7\textwidth]{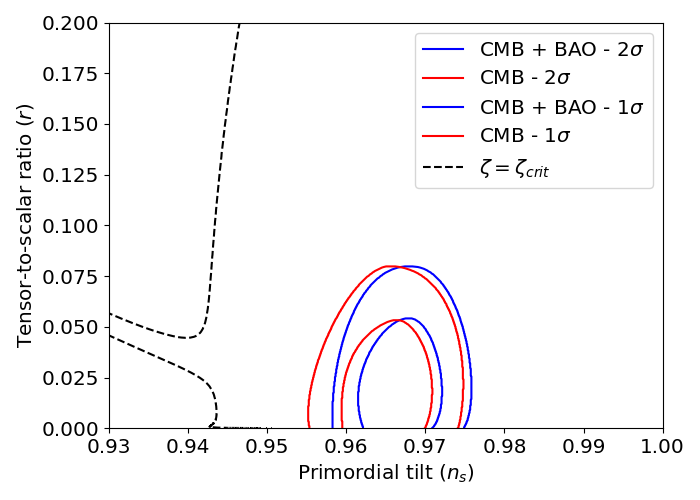}
\caption{Black dashed lines show prediction of $n_s$ and $r$ for parameters giving exactly one point where $\partial V_{\rm eff}/\partial\phi = 0$ for $n=2$ (top) and $n=4$ (bottom), computed using the full equations. For parameters whose predictions are located to the left of the black dashed line the model has two $\phi_{\rm crit}$, while the region to the right is devoid of them. Together shown by red and blue contours are the Planck result~\cite{Planck:2018jri}.}
\label{figPhiCritBound2}
\end{figure}

Finally, let us mention that for the monomial potential models this whole discussion becomes purely theoretical. For the range of parameters needed to obtain results consistent with the Planck data, it turns out that \eqref{PhiCritPolyCondition} is satisfied. The boundary between parameter range which possess $\phi_{\rm crit}$ and which does not is shown in Fig.~\ref{figPhiCritBound2}. This rules out any ultra-slow-roll epoch in GB inflation with a dilaton-like coupling and quadratic or quartic potentials. 

\section{Monomial potential models}
\label{sec:mono}

\begin{figure}[h]
\centering
\includegraphics[width=.7\textwidth]{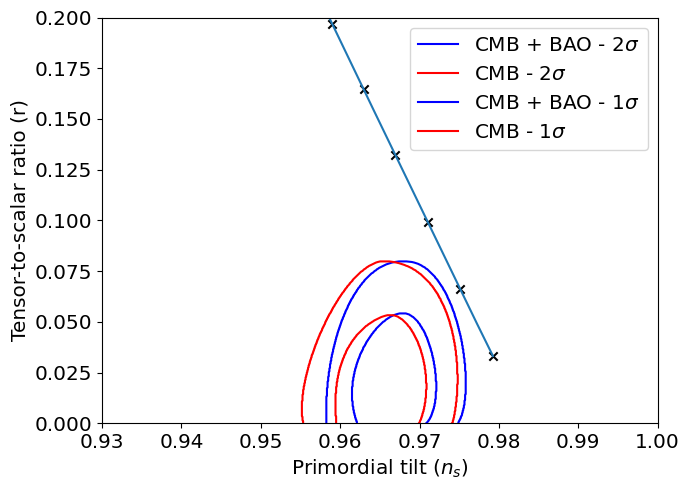}
\caption{Prediction of $n_s$ and $r$ for the polynomial potential model with various values of $n$. The values of $n$ range from 0.5 (bottom) to 3 (top). The crosses mark an increase in $n$ by 0.5.}
\label{figPolyVanilla}
\end{figure}

In this section we will discuss the phenomenological predictions for monomial potentials of the form 
\begin{equation}
\label{PolynomialPotentialEQ}
    V=g\phi^n.
\end{equation}
This includes the simplest inflation model with $n=2$. In pure general relativity, the polynomial potential models are ruled out by the Planck data~\cite{Planck:2018jri} as shown in Fig.~\ref{figPolyVanilla}. On the other hand, the presence of the GB term can save them as will be shown here.

For the calculation of $n_s$ and $r$ in the slow-roll approximation it is more suitable to parametrize the model as
\begin{equation}
    V=\frac{3 \phi^{n} \zeta}{4 \xi_{0}},
\end{equation}
which results in $\xi_0$ dropping out of the equations. Given a value of $\zeta$ we can obtain $\xi_0$ by fixing the amplitude of the scalar perturbation. The standard slow-roll parameters are
\begin{equation}
    \epsilon = \frac{n^{2}}{2 \phi^{2}},~~~\eta = \frac{n \left(n - 1\right)}{\phi^{2}}.
\end{equation}
The new ones related to the GB coupling become
\begin{equation}
    \alpha = - \frac{3 \lambda \phi^{n - 1} \zeta n e^{- \lambda \phi}}{2},~~~\beta = \lambda^{2} \phi^{n} \zeta e^{- \lambda \phi},~~~\gamma = 2 \lambda^{2} \phi^{2 n} \zeta^{2} e^{- 2 \lambda \phi}.
\end{equation}
The field equation for $\phi$ becomes
\begin{equation}
    \frac{d\phi}{dN}=\lambda \zeta  \phi^{n} e^{- \lambda \phi} - \frac{n}{\phi}.
\end{equation}
To ensure the validity of the slow-roll approximation, the behavior of $\alpha$, $\beta$ and $\gamma$ should be reasonable. In order for them to be bounded we require $\lambda>0$. In the opposite case the GB term would make $V_{\rm eff}$ steeper. $\alpha$ has a minimum at
\begin{equation}
\label{eq:AlphaExtreme}
    \phi_0=\frac{n-1}{\lambda},~~~\alpha_0=- \frac{3 \lambda \zeta n \left(\frac{n - 1}{\lambda}\right)^{n - 1} e^{1 - n}}{2}.
\end{equation}
For $\beta$ we have
\begin{equation}
\label{eq:BetaExtreme}
    \phi_0=\frac{n}{\lambda},~~~\beta_0=\lambda^{2} \zeta \left(\frac{n}{e \lambda}\right)^{n},
\end{equation}
and finally for $\gamma$ we have
\begin{equation}
\label{eq:GammaExtreme}
    \phi_0=\frac{n}{\lambda},~~~\gamma_0=2 \lambda^{2} \zeta^{2} \left(\frac{n}{\lambda}\right)^{2 n} e^{- 2 n}.
\end{equation}

\subsection{Quadratic potential}

\subsubsection{Scalar spectral index and tensor-to-scalar ratio}

\begin{figure}[H]
\centering
\includegraphics[width=.7\textwidth]{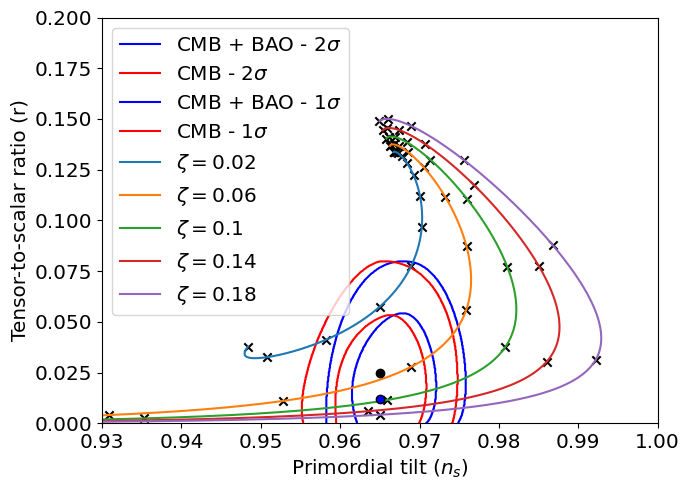}
\includegraphics[width=.7\textwidth]{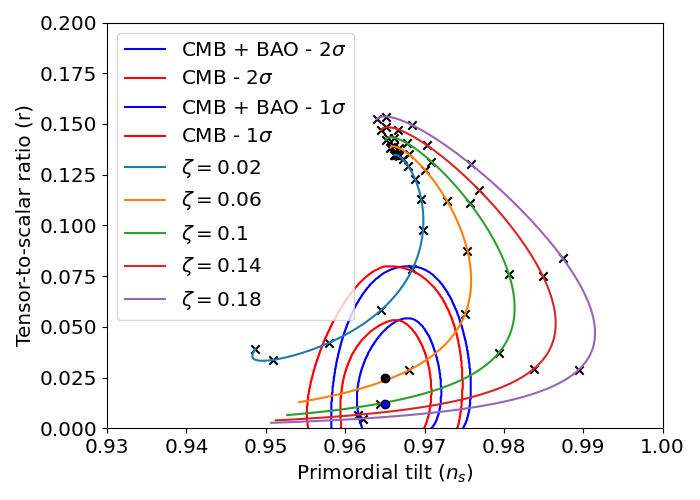}
\caption{Prediction of $n_s$ and $r$ in the quadratic potential for various values of $\zeta$ and $\lambda$ computed using the slow-roll approximation (top) and using the the full equations of motion with a reference value of $g=2\cdot 10^{-10}$ (bottom). 
The value of $\lambda$ varies along each line from 0.04 to 0.6. The crosses mark an increase in $\lambda$ by 0.04. Two dots represent our reference parameter points.}
\label{figPolyN2}
\end{figure}

The simplest choice $V=\frac{1}{2}m^2\phi^2$ can be made consistent with the Planck data using the GB term. We have computed the predictions for $n_s$ and $r$ using the slow-roll approximation in Ref.~\cite{Mudrunka:2023wxy}. 
For other choices of the GB coupling function $\xi$ see Ref.~\cite{Jiang:2013gza}. Here we verify the precision of the approximation by solving the full equation of motion and the full Friedmann equation. Using the slow-roll approximation we obtain the prediction for $n_s$ and $r$ as shown in the top panel of Fig.~\ref{figPolyN2}.

The validity of the slow-roll approximation can be apriori estimated using \eqref{eq:AlphaExtreme}, \eqref{eq:BetaExtreme} and \eqref{eq:GammaExtreme}. For $n=2$ we have
\begin{equation}
    \alpha_0=-\frac{3\zeta}{e},~~~\beta_0=\frac{4\zeta}{e^2},~~~\gamma=\frac{32\zeta^2}{e^4\lambda^2}\approx 0.59\left(\frac{\zeta}{\lambda}\right)^2.
\end{equation}
Note that to obtain large $\gamma_0$ we would need $\zeta/\lambda\geq 1$, which lies in the region affected by the initial condition fine tuning problem. The validity of the slow-roll approximation is directly confirmed by solving the full equations of motion as presented in the bottom panel of Fig.~\ref{figPolyN2}.
In this plot we excluded the points affected by the initial condition fine tuning. 

\subsubsection{Reheating GW spectra}

As explored in Refs.~\cite{Ema:2021fdz,Mudrunka:2023wxy} and also shown in App.~\ref{sec:decay}, the GB term induces the inflaton decay into the graviton pair, resulting in the GW background in the present universe.
In order to make predictions for the GW spectra, we choose two points of interest that lie within the $1\sigma$ confidence regions of the Planck data. First is marked by a black dot and is given by $(n_s=0.965, r=0.0247)$, the second is marked by a blue dot and corresponds to a lower tensor-to-scalar ratio $(n_s=0.965, r=0.01)$. The black point corresponds to $\lambda=0.218$ and $\zeta=0.0532$. The maxima of the absolute values of the slow-roll parameters during the inflation are $\alpha_m=0.0587$, $\beta_m=0.0288$ and $\gamma_m=0.0349$ is this case. The blue point corresponds to $\lambda=0.286$ and $\zeta=0.106$. The maxima of the absolute values of the slow-roll parameters are $\alpha_m=0.117$, $\beta_m=0.0574$ and $\gamma_m=0.0917$ for this case. 

As also shown in App.~\ref{sec:decay}, we obtain lower bound on the reheating temperature to avoid too much dark radiation as $\sim 0.4\cdot10^{8}$\,GeV for the blue point and $\sim 0.2\cdot10^8$\,GeV
for the black point. Generally getting a lower tensor-to-scalar ratio requires stronger coupling to the GB term which leads to higher decay rates and higher minimal reheating temperatures.
\begin{figure}[H]
\label{figPolyGWSpectrum}
\centering
\includegraphics[width=.7\textwidth]{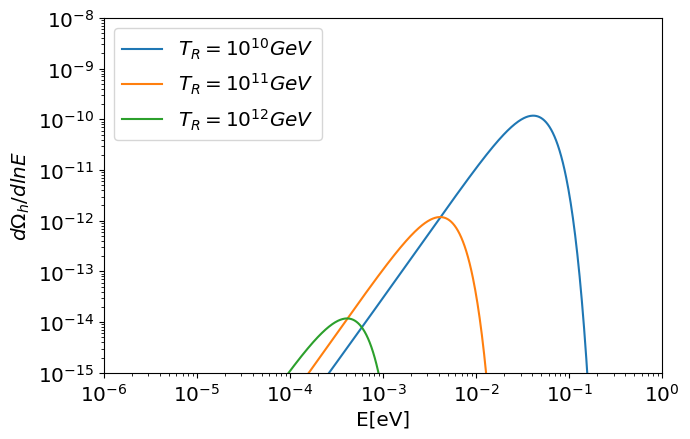}
\includegraphics[width=.7\textwidth]{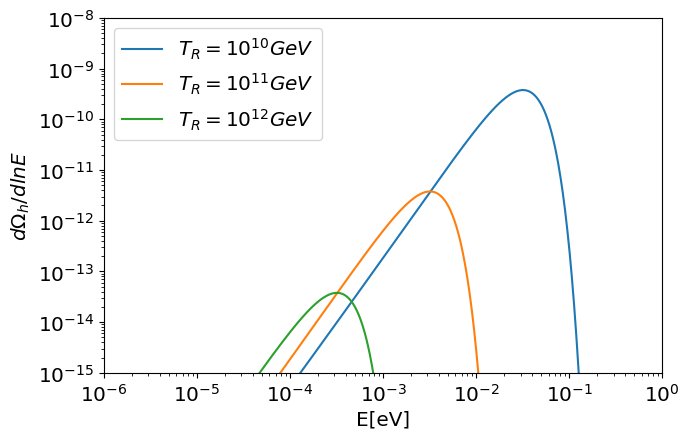}
\caption{The GW spectra resulting from the decay induced by the GB term for the points $n_s=0.965$, $r=0.0247$ (left figure) and $n_s=0.965$, $r=0.01$ (right figure) and $n=2$.}
\end{figure}

\subsection{Quartic potential}

\begin{figure}[H]
\centering
\includegraphics[width=.7\textwidth]{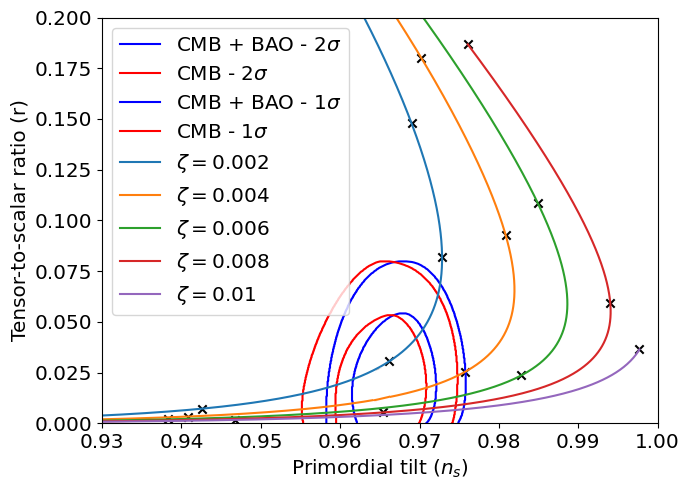}
\includegraphics[width=.7\textwidth]{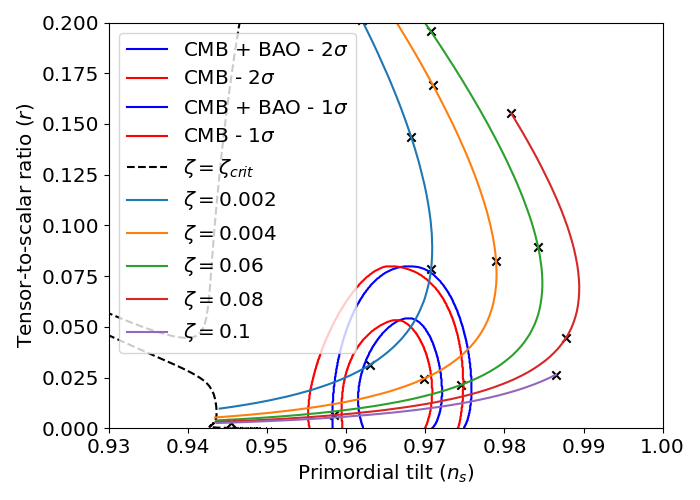}
\caption{Prediction of $n_s$ and $r$ in the quartic potential model for various values of $\zeta$ and $\lambda$ computed using the slow-roll approximation (top) and using the full equations of motion with a reference value of $g=2\cdot10^{-11}$ (bottom). The value of $\lambda$ varies along each line from 0.3 to 0.5. The crosses mark an increase in $\lambda$ by 0.02. See Fig.~\ref{figPhiCritBound2} for the black-dashed line.
}
\label{figPolyN4}
\end{figure}

In the case $n=4$ the necessary correction from the GB term is much larger compared to the $n=2$ case. This causes the slow-roll parameters to reach higher magnitudes, and as will be shown here, this makes the prediction lose its precision. The predictions for $n_s$ and $r$ computed using the slow-roll approximation are presented in the top panel of Fig.~\ref{figPolyN4}.

The slow-roll results are quite imprecise. The same results computed using the full equations of motion and not expanding the Hankel function arguments to first order in the slow-roll parameters are presented in the bottom panel of Fig.~\ref{figPolyN4}. 

\begin{figure}[H]
\centering
\includegraphics[width=.48\textwidth]{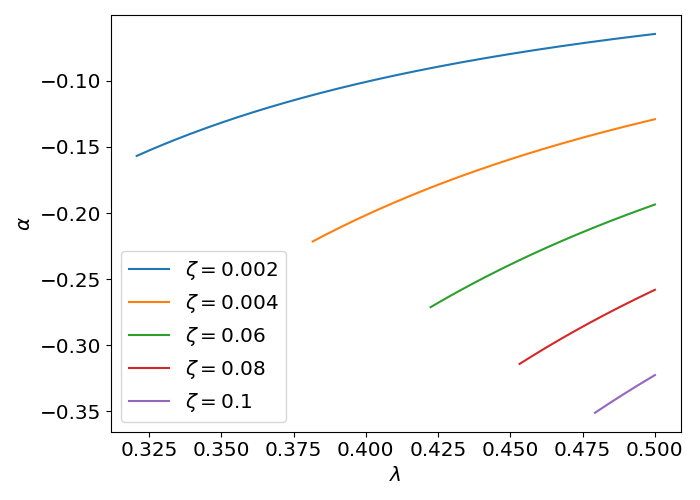}
\includegraphics[width=.48\textwidth]{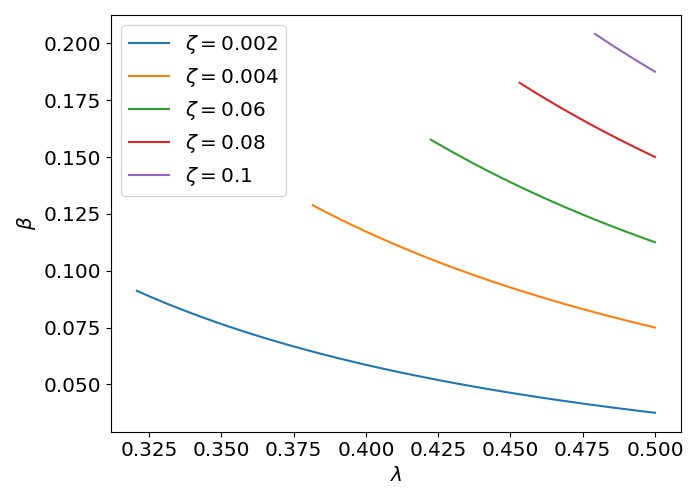}
\includegraphics[width=.48\textwidth]{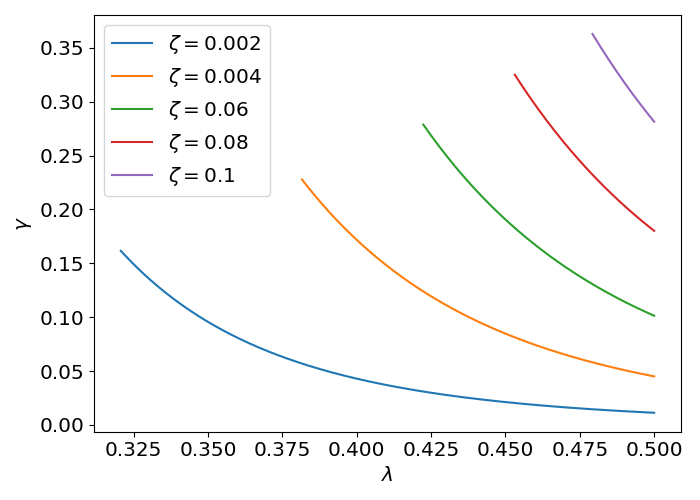}
\caption{Minimal values of $\alpha$ and maximal values $\beta$ and $\gamma$ between end of inflation and horizon exit at 60 e-folds for $n=4$ and a range of $\lambda$ and $\zeta$. Only the results for the range where $\frac{\partial V_{\rm eff}}{\partial\phi}>0$ everywhere are shown.}
\label{figAlphaVals4}
\end{figure}

\begin{figure}[H]
\centering
\includegraphics[width=.6\textwidth]{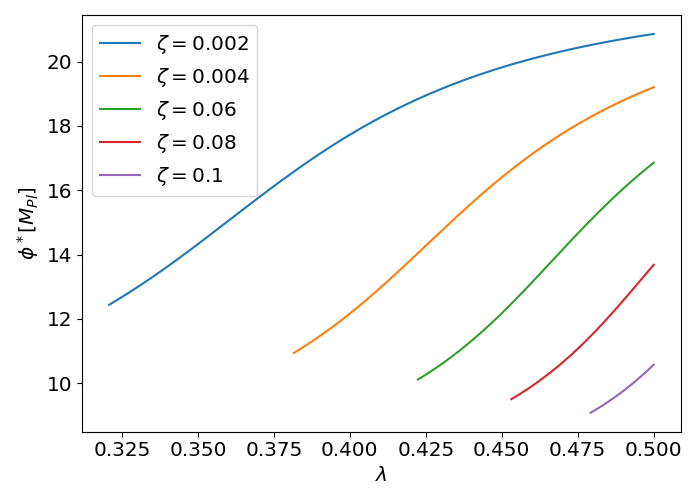}
\caption{Values of $\phi$ at horizon exit at 60 e-folds for $n=4$ and a range of $\lambda$ and $\zeta$. Only the results for the range where $\frac{\partial V_{\rm eff}}{\partial\phi}>0$ everywhere are shown.}
\label{figPhiVals4}
\end{figure}

We see that the slow-roll approximation together with the formulas that use first order expansions overestimates $n_s$ and underestimates $r$. To illustrate how much the slow-roll approximation fails we present plots of the maximal values of $\vert \alpha\vert$, $\vert \beta\vert$, $\vert \gamma\vert$ in Fig.~\ref{figAlphaVals4}. 
$\alpha$, $\beta$ and $\gamma$ reach their extrema at
\begin{equation}
    \phi_{\alpha}=\frac{3}{\lambda},~~~\phi_{\beta,\gamma}=\frac{4}{\lambda},
\end{equation}
which as illustrated by Fig.~\ref{figPhiVals4} happens after the horizon exit. This results in a non-negligible error in the resulting value of $\phi$ at the horizon exit.

In order to use \eqref{eqDecayRate} to compute the GW spectrum produced during reheating, we need to substitute the effective ``mass'' of the inflaton, $m^2 \sim g \phi^2$, but it is highly suppressed at the later stage of reheating.
If a small mass term for $\phi$ is introduced, it may significantly enhance the GW signal. This term would, however, become a new free parameter and the model would lose its predictive power.  

\section{Generalized dilaton-like coupling}
\label{sec:general}

In this section we will study a generalization of the dilaton-like coupling model given by
\begin{equation}
    \xi=8\xi_0 e^{-\lambda\phi^m/M_{\rm pl}^m},~~~m\in \mathbb{N}.
\end{equation}
An interesting class of models is formed by those with even $m$, resulting in a $Z_2$ symmetry. 
We consider models with a monomial potential given by
\begin{equation}
    V=g\phi^n.
\end{equation}
For the slow roll calculation it is more suitable to parametrize the potential as
\begin{equation}
    V=\frac{3 \zeta}{4 \xi_{0}}\phi^{n} .
\end{equation}
The slow roll field equation becomes
\begin{equation}
\label{eq:DilatonGeneralMonomialSlowRollEq}
    \frac{d\phi}{dN}= \lambda \phi^{m + n - 1} \zeta n e^{- \lambda \phi^{m}} - \frac{n}{\phi},
\end{equation}
and the slow roll parameters become
\begin{align}
    &\epsilon=\frac{n^{2}}{2 \phi^{2}},~~~\eta=\frac{n \left(n - 1\right)}{\phi^{2}},~~~\alpha=- \frac{3 \lambda \phi^{m + n - 2} \zeta m n e^{- \lambda \phi^{m}}}{2},\\
    &\beta=\lambda \phi^{m + n - 2} \zeta m \left(\lambda \phi^{m} m - m + 1\right) e^{- \lambda \phi^{m}},~~~\gamma=2 \lambda^{2} \phi^{2 m + 2 n - 2} \zeta^{2} m^{2} e^{- 2 \lambda \phi^{m}}.
\end{align}

\begin{figure}[H]
\begin{center}
   \includegraphics[width=8.15cm]{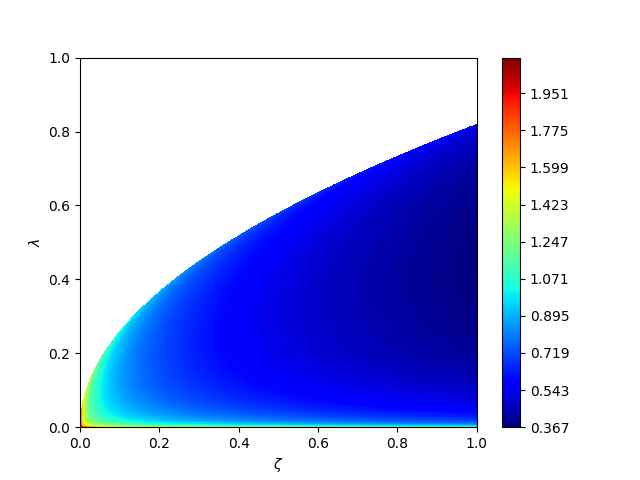}
   \includegraphics[width=8.15cm]{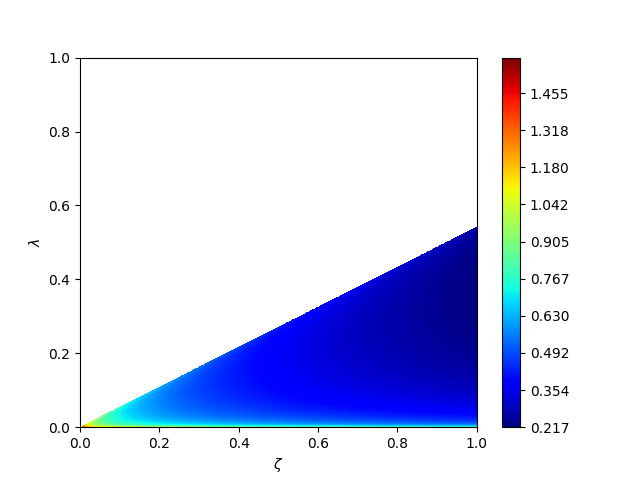}
   \includegraphics[width=8.15cm]{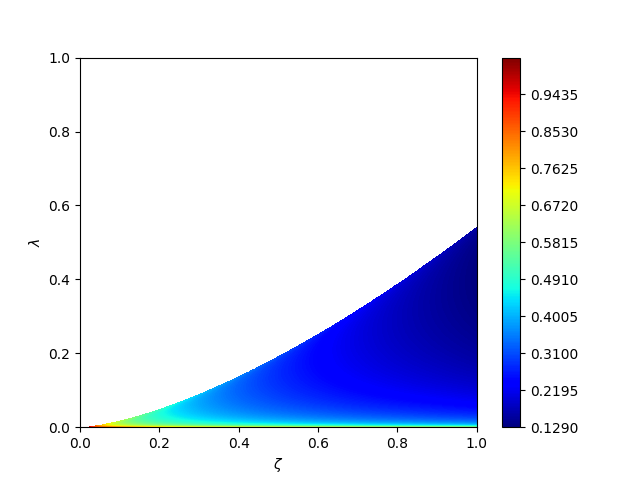}
   \includegraphics[width=8.15cm]{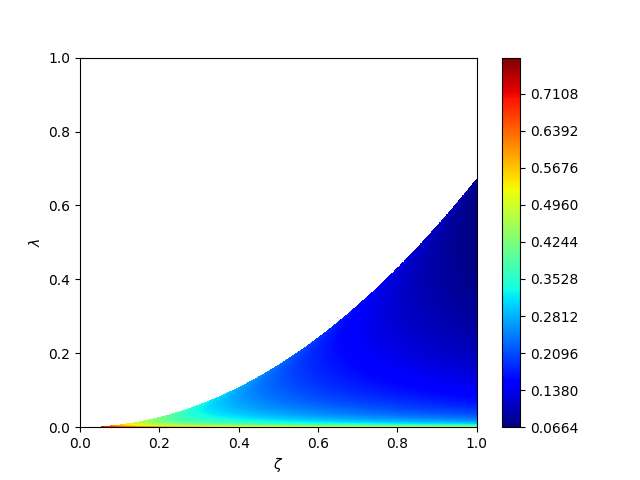}
  \end{center}
  \caption{Plot of $\log_{10}\phi_{\rm crit}$ corresponding to the $W_0$ branch of the Lambert W function. The white areas correspond to parameters where no real solution for $\phi_{\rm crit}$ exists. The potential and coupling function parameters are $n=2,~m=1$ (top left), $n=2,~m=2$ (top right), $n=2,~m=3$ (bottom left) and $n=2,~m=4$ (bottom right).
 }
  \label{fig:critLevels}
\end{figure}

To compute the critical levels where $\frac{\partial V_{\rm eff}}{\partial\phi}=0$, we need to solve an equation of the form $x^Ne^{Bx}=A$. A solution in terms of the Lambert W function\ can be found using the factorization
\begin{equation}
    \left(\frac{Bx}{N}\right)^N\left(e^{\frac{Bx}{N}}\right)^N=A\left(\frac{B}{N}\right)^N
\end{equation}
and is given by
\begin{equation}
    x=\frac{N}{B}W\left(\frac{B}{N}A^{1/N}\right).
\end{equation}
Applying this formula to \eqref{eq:DilatonGeneralMonomialSlowRollEq} we get
\begin{equation}
    \phi_{\rm crit}=\left[-\frac{m+n}{\lambda m}W\left(-\frac{\lambda m}{m+n}\left(\frac{n}{\lambda\zeta m}\right)^{\frac{m}{m+n}}\right)\right]^{\frac{1}{m}},
\end{equation}
where in principle one should consider all possible complex square roots and branches of the W function, but of course we are interested only in real solutions. The W function is has two real branches for arguments greater than or equal to $-\frac{1}{e}$, so in order to have the possibility of being trapped in a local minimum of the effective potential we need
\begin{equation}
    \zeta>\frac{n}{m}\left(\frac{em}{m+n}\right)^{\frac{m+n}{m}}\lambda^{\frac{m+n}{m}-1},
\end{equation}
a generalization of the result \eqref{PhiCritPolyCondition}. The values of $\phi_{\rm crit}$ corresponding to the $W_0$ branch are shown in Fig.~\ref{fig:critLevels}.

Similarly to the $m=1$ case, the models are compatible with the Planck data. Because of the additional factor of $\phi^{m-1}$ in the derivative of the coupling, the effect of the GB term becomes much stronger and therefore very low values of $\lambda$ and $\zeta$ are required. We present the predictions for $m=2$ and $n=2,3,4$ computed using the slow roll approximation.
\begin{figure}[H]
\centering
\includegraphics[width=.7\textwidth]{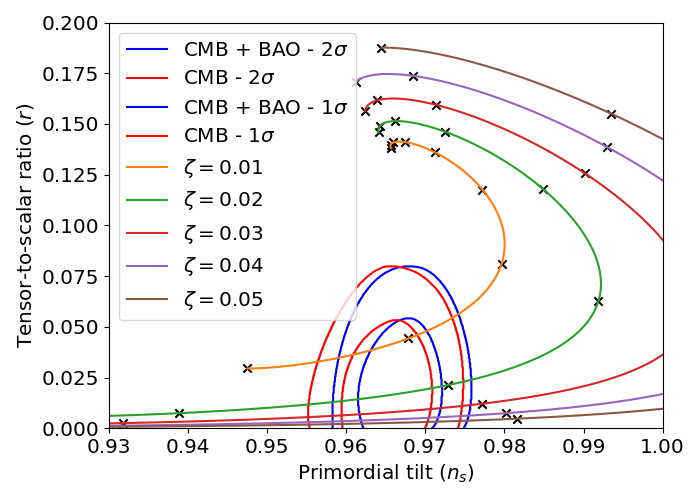}
\caption{Prediction of $n_s$ and $r$ in the quadratic potential model with $m=2$ for various values of $\zeta$ and $\lambda$ computed using the slow-roll approximation. The value of $\lambda$ goes from 0.005 to 0.045 along each line. Black crosses mark an increase in $\lambda$ by 0.005.}
\label{figGenMono2}
\end{figure}
\begin{figure}[H]
\centering
\includegraphics[width=.7\textwidth]{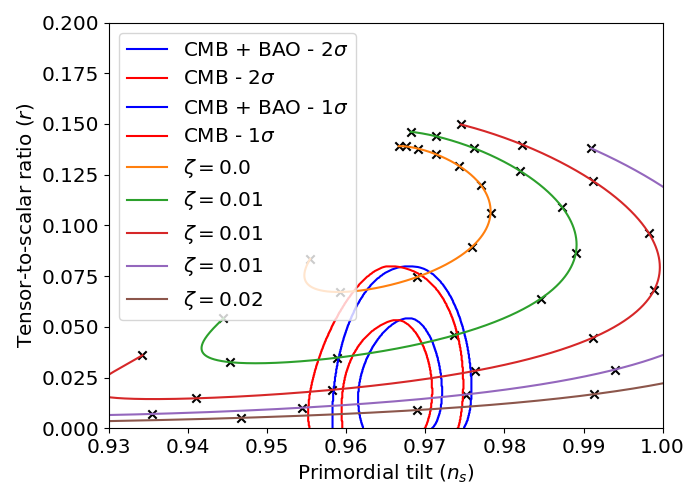}
\caption{Prediction of $n_s$ and $r$ in the quadratic potential model with $m=3$ for various values of $\zeta$ and $\lambda$ computed using the slow-roll approximation. The value of $\lambda$ goes from 0.0001 to 0.0021 along each line. Black crosses mark an increase in $\lambda$ by 0.0002.}
\label{figGenMono2}
\end{figure}
\begin{figure}[H]
\centering
\includegraphics[width=.7\textwidth]{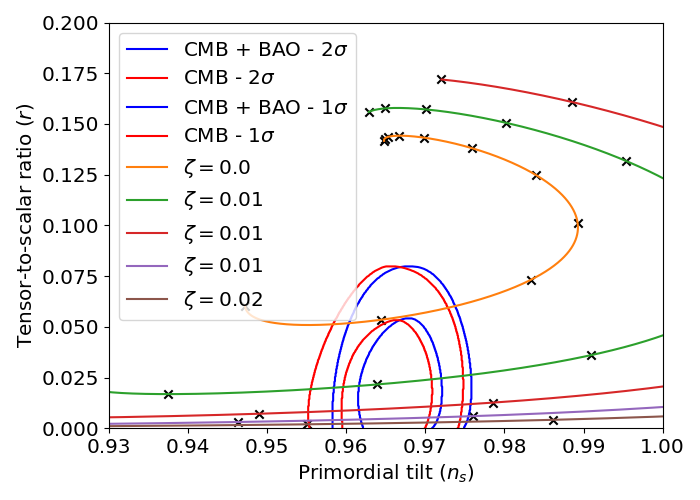}
\caption{Prediction of $n_s$ and $r$ in the quadratic potential model with $m=4$ for various values of $\zeta$ and $\lambda$ computed using the slow-roll approximation. The value of $\lambda$ goes from 0.00001 to 0.00021 along each line. Black crosses mark an increase in $\lambda$ by 0.00002.}
\label{figGenMono2}
\end{figure}
Unlike the $m=1$ case, the reheating GW spectrum induced by the GB coupling is going to be heavily suppressed in models with $n\geq2$ due to the additional factors of $M_{\rm pl}^{-1}$.

\section{Conclusion}

We studied the validity of the slow-roll approximation presented in Ref.~\cite{Satoh:2008ck} for monomial potential inflation models with the GB term and a dilaton-like coupling function. Such models are theoretically motivated by string theory corrections to gravity. In the quadratic potential case the model is compatible with the most recent Planck data constraints on the values of the scalar perturbation spectral index and tensor-to-scalar ratio. The model also predicts a unique footprint in the GW spectrum produced by decay of inflaton particles into graviton pairs during the reheating epoch.

We have shown that in monomial potential models there exists a minimal negative initial value of $\dot{\phi}$ below which the Friedmann equations do not admit any positive solutions for $H$. General analytic formula for this condition has been obtained. We also analytically described the existence of two critical values of the inflaton field, between which the model exhibits the existence of a local minimum in its effective potential, possibly leading to everlasting inflation. These phenomena constrain the possible initial conditions leading to viable inflation followed by a reheating epoch. It turns out that in the case of quadratic and quartic potentials the conditions do not affect the range of parameters required for compatibility with the Planck data.

The precision of the slow-roll formulas has been verified for the case of a quadratic potential. In the case of a quartic potential the formulas become quite imprecise. This is a result of a stronger effect of the GB term required to reach the low tensor-to-scalar ratios favored by the Planck data. We numerically solved the full equations of motion, compared the results and discovered a non-negligible difference in the predicted values of $n_s$ and $r$. A more precise analysis would also involve solving the Mukhanov-Sasaki equation numerically for a large number of modes instead of using the first order Hankel function formulas, which was not done here.

The existence of the critical values of the inflaton field where the derivative of the effective potential of the model becomes zero points at the possibility of an ultra-slow-roll epoch resulting in formation of primordial black holes.

\section*{Acknowledgements}
This work was supported by JST, the establishment of university fellowships towards the creation of science technology innovation, Grant Number JPMJFS2102.
This work was supported by JSPS KAKENHI Grant Numbers 24K07010 [KN].
This work was supported by World Premier International Research Center Initiative (WPI), MEXT, Japan.

\appendix

\section{Inflaton decay through the Gauss-Bonnet term}
\label{sec:decay}

The coupling between inflaton field and the GB term creates new pathways through which decay into gravitons occurs~\cite{Ema:2021fdz}, leading to a signature footprint in the spectrum of GWs produced during the reheating phase~\cite{Mudrunka:2023wxy}.\footnote{
    See also Refs.~\cite{Koshelev:2022wqj,Tokareva:2023mrt,Landini:2025jgj} for other cosmological applications of particle decay into the graviton pair.
} Expanding the coupling function around the minimum of the inflaton potential, which we consider to be $\phi_0=0$ in this section, we get an infinite series of interaction terms 
\begin{equation}
    \mathcal{L}_{\rm GB-\phi}=\sum_{n=1}^{\infty}\frac{\xi^{(n)}(0)}{n!}\phi^n\sqrt{-g}\left(R_{\mu\nu\alpha\beta}R^{\mu\nu\alpha\beta}-4R_{\mu\nu}R^{\mu\nu}+R^2\right). 
\end{equation}
In the late phase of reheating when $\phi\approx 0$ the gravitational field equations reduce to those of standard GR and thus for on-shell gravitons we have $R_{\mu\nu}=R=0$, leaving only the Kretschmann scalar as a source of the interaction. The decay into one graviton $(n\phi\rightarrow h)$ is unimportant, because the Riemann tensor is proportional to $H$ and $\dot{H}$, which are also small for small values of $\phi$, as dictated by the Friedmann equations. For decay into graviton pairs $(n\phi\rightarrow hh)$ we again have terms proportional to the unperturbed Riemann tensor, but also 
\begin{equation}
    I=\sqrt{-g}\delta R^{\mu\nu\alpha\beta}\delta R_{\mu\nu\alpha\beta},
\end{equation}
where $\delta R^{\mu\nu\alpha\beta}$ is the first order perturbation of the Riemann tensor. By expanding the covariant derivatives in $\delta R^{\mu\nu\alpha\beta}$ and separating the Christoffel symbols, which are proportional to $H$, we come to the conclusion that the dominant term which drives the decay comes from the expansion of the Riemann tensor around flat Minkowski spacetime. The resulting interaction Lagrangian is (we take $a\approx 1$)
\begin{equation}
\begin{split}
    \mathcal{L}_{\rm GB-\phi}=\sum_{n=1}^{\infty}\frac{\xi^{(n)}(0)}{n!}\frac{\phi^n}{4}\left(\partial^\alpha\partial^\nu h^{\mu\beta}+\partial^\beta\partial^\mu h^{\nu\alpha}-\partial^\alpha\partial^\mu h^{\nu\beta}-\partial^\beta\partial^\nu h^{\mu\alpha}\right) \\ \times \left(\partial_\alpha\partial_\nu h_{\mu\beta}+\partial_\beta\partial_\mu h_{\nu\alpha}-\partial_\alpha\partial_\mu h_{\nu\beta}-\partial_\beta\partial_\nu h_{\mu\alpha}\right).
\end{split}   
\end{equation}
For the graviton 4-momentum $k$ and $k'$, the resulting vertex is given by
\begin{equation}
    V(k,k^{\prime})_{\mu\nu\alpha\beta}=\kappa\left[(k\cdot k^{\prime})^2\eta_{\mu\alpha}\eta_{\nu\beta}-\eta_{\mu\alpha}k_\nu k^{\prime}_\beta(k\cdot k^{\prime})-\eta_{\nu\beta}k_\alpha k^{\prime}_\mu(k\cdot k^{\prime})+k_\mu k_\nu k^{\prime}_\alpha k^{\prime}_\beta\right],
\end{equation}
where $\kappa$ is the corresponding coupling constant obtained from the Taylor series of $\xi$. Including a symmetry factor of 2, multiplying by the polarization vectors and squaring, we obtain the transition amplitudes
\begin{equation}
    \vert \mathcal{A} \vert_{++}^2=64\kappa^2(k \cdot k^{\prime})^4,~~~\vert \mathcal{A} \vert_{\times\times}^2=64\kappa^2(k \cdot k^{\prime})^4,~~~\vert \mathcal{A} \vert_{+\times}^2=\vert \mathcal{A} \vert_{\times+}^2=0,
\end{equation}
where $+$ and $\times$ denote the two graviton polarizations and we employ the transverse-traceless gauge $h^{\mu}_\mu=k_\mu h^\mu_\nu=0$. The resulting decay rate for $\phi\rightarrow hh$ is then~\cite{Ema:2021fdz}
\begin{equation}
    \label{eqDecayRate}
    \Gamma_{\phi\rightarrow hh}=\frac{\kappa^2 m^7}{4\pi M_{\rm pl}^4}.
\end{equation}
In principle we should also take into account decay into three and more gravitons, but with each interacting graviton the amplitude is suppressed by and additional factor of $1/M_{\rm pl}^2$, making these decay paths negligible.

The decay-produced gravitons form dark radiation and contribute to the effective number of extra neutrino species $\Delta N_{\rm eff}$, which is constrained by the CMB observation~\cite{Planck:2018vyg}.
To avoid too much dark radiation, the branching ratio of the inflaton decay into the graviton pair must be subdominant.
The total decay width into the Standard Model particles is parametrized by the reheating temperature $T_{\rm R}$ as
\begin{equation}
    \Gamma_{\rm tot}=\sqrt{\frac{\pi^2 g_*(T_{\rm R})}{90}\frac{T_{\rm R}^4}{M_{\rm pl}^2}}.
\end{equation}
The effective number of extra neutrino species is given by
\begin{equation}
    \Delta N_{\rm eff} \simeq \frac{43}{7}\left(\frac{10.75}{g_*(T_{\rm R})}\right)^{1/3} \frac{ \Gamma_{\phi\rightarrow hh}}{\Gamma_{\rm tot}}.
\end{equation}
Thus, for fixed inflaton mass and GB couplings, we obtain lower bound on $T_{\rm R}$ to avoid too much dark radiation.


\end{document}